\documentclass[journal]{new-aiaa}

\usepackage[utf8]{inputenc}
\usepackage{graphicx}
\usepackage[version=4]{mhchem}
\usepackage{siunitx}
\usepackage{longtable,tabularx}
\usepackage{float}
\usepackage{setspace}
\usepackage[dvips]{epsfig}
\usepackage{bm}
\usepackage{epstopdf}
\usepackage{subfig}
\usepackage{algorithm,algpseudocode}
\usepackage{color, soul}
\usepackage{tabularx}
\usepackage{booktabs}
\usepackage{caption}

\newcolumntype{Y}{>{\raggedleft\arraybackslash}X}

\makeatletter

\newcommand{\bbm}{\begin{bmatrix}}
\newcommand{\ebm}{\end{bmatrix}}
\newcommand{\be}{\begin{equation}}
\newcommand{\ee}{\end{equation}}
\newcommand{\bea}{\begin{eqnarray}}
\newcommand{\eea}{\end{eqnarray}}
\newcommand{\beai}{\begin{IEEEeqnarray}}
\newcommand{\eeai}{\end{IEEEeqnarray}}
\newcommand{\eeais}{\end{IEEEeqnarray*}}
\newcommand{\beas}{\begin{eqnarray*}}
\newcommand{\eeas}{\end{eqnarray*}}

\usepackage{xcolor}
\hypersetup{
    colorlinks,
    linkcolor={red!50!black},
    citecolor={blue!50!black},
    urlcolor={blue!80!black}
}

\makeatother

\begin{document}
\title{Model-Predictive Planning and Airspeed Regulation to Minimize Flight Energy Consumption\footnote{This manuscript originated from the conference presentation in AIAA SciTech on January 23, 2023, at National Harbor, Maryland, USA. AIAA paper number: AIAA 2023-0303.}}

\author{Trevor Karpinski \footnote{Master's student, Department of Mechanical and Aerospace Engineering.}}
\affil{New Mexico State University, Las Cruces, New Mexico, 88003, USA}

\author{Alexander Blakesley\footnote{PhD student, Department of Civil, Environmental and Geomatic Engineering.}, Jakub Krol\footnote{Research Fellow, Department of Civil, Environmental and Geomatic Engineering.}, Bani Anvari\footnote{Lecturer, Department of Civil, Environmental and Geomatic Engineering.}}
\affil{Centre for Transport Studies, CEGE, University College London, Gower Street, WC1E 6BT, United Kingdom.}

\author{George Gorospe\footnote{Senior Research Scientist}} 
\affil{NASA Ames Research Center, Moffett Field, CA, 94035, USA}

\author{Liang Sun\footnote{Associate Professor, Department of Mechanical Engineering. AIAA Senior Member.}}
\affil{Baylor University, Waco, Texas, 76798, USA}

\date{}
\maketitle
\begin{abstract}
Although battery technology has advanced tremendously over the past decade, it continues to be a bottleneck for the mass adoption of electric aircraft in long-haul cargo and passenger delivery.  The onboard energy is expected to be utilized in an efficient manner. Energy concumption modeling research offers increasingly accurate mathematical models, but there is scant research pertaining to real-time energy optimization at an operational level. Additionally, few publications include landing and take-off energy demands in their governing models. This work presents fundamental energy equations and proposes a proportional-integral-derivative (PID) controller. The proposed method demonstrates a unique approach to an energy consumption model that tracks real-time energy optimization along a predetermined path. The proposed PID controller was tested in simulation, and the results show its effectiveness and accuracy in driving the actual airspeed to converge to the optimal velocity without knowing the system dynamics. We also propose a model-predictive method to minimize the energy usage in landing and take-off by optimizing the flight trajectory..
\end{abstract}

\section{Introduction\label{sec:Introduction}}
An unmanned aerial system (UAS) consists of an unmanned aerial vehicle and its supporting devices, typically autonomous, set out to complete a specified task. 
Multicopter UASs, also referred to as drones, have great utility due to their ability to fly over geographical barriers and have been utilized in various applications such as package delivery in urban environments~\cite{Arezoo2021}, search and rescue missions~\cite{MISHRA2020}, or combating building fires~\cite{estrada2021trajectory}.
Nonetheless, there are unique considerations that need to be recognized to determine the comparative effectiveness over ground-based vehicles. 
For example, battery storage technology severely constrains the performance of both automotive and multicopter modes.
Furthermore, realistic optimization of energy expenditure is made difficult by the complicated nature of multicopter dynamics.
While modern literature covers energy expenditure estimation for drones, or path planning optimization in the Traveling Salesman problem, there appears to be little to no research with both fields in mind. 
Thus, long-term offline path planning that focuses on energy optimization combined with a real-time motion controller could provide a unique approach to modern problems of UAS applications. With these considerations, the energy-aware path planning and control of a drone seeks to greatly increase the utility and application of drones that would allow for greater mass adoption of UASs.
Existing research for analysis of multicopter energy expenditure can be divided into two areas: estimation of energy usage, and energy optimization through path planning. 
The first area of research - calculating the energy expenditure of a drone over a time interval or at a specific point in time - is non-trivial and requires an in-depth analysis of the aerodynamic properties of a drone. 
Zhang et al.~\cite{zhang2021energy} conducted a meta-analysis on various energy consumption models, such as the ones developed by Kirchstein~\cite{KIRSCHSTEIN2020102209}, Dorling et al.~\cite{Dorling2017}, D'Andrea~\cite{D'Andrea2014}, and Liu et al.~\cite{Liu2017}, and categorized two approaches for energy expenditure calculation: those that assume lift-to-drag ratio to be the dominant factor, and those assuming that drones in forward flight will have similar energy expenditure formulation as a hovering helicopter.  

D'Andrea~\cite{D'Andrea2014} falls under a model for energy expenditure that solely focuses on a critical parameter: the lift-to-drag ratio. It stems from a general-purpose analysis that seeks to approximate many classes of drones under the singular model, thus avoiding the lengthy analysis pertaining to each class of drone. 
An alternative integrated model was introduced in Figliozzi~\cite{Figliozzi2017} with an aim to analyze energy expenditure for the purpose of approximating greenhouse gas emissions for environmental impact. While these models serve an important role, their general formulation may not be accurate enough for optimization purposes.

Models introduced by Kirchstein~\cite{KIRSCHSTEIN2020102209} and Dorling et al.~\cite{Dorling2017} fall under the category of component models, which are models with the base assumption that power requirements for a helicopter in forward flight, ascent, and descent are approximately equal to the power requirements while hovering. Liu et al.~\cite{Liu2017} assessed the accuracy of a component model by conducting field tests that demonstrate a 16.5\% error from the theoretical calculations that under-estimates the real-world energy demands for forward flight. The component models provide a more comprehensive analysis of the dynamics of multicopters, and thus more accurately represent the energy demands of the UAS, but at the cost of being heavily reliant on dynamic properties of a specific drone, such as the coefficient of drag and blade-lift coefficient. After careful evaluation of the presented literature, the energy model formulated by Kirchstein~\cite{KIRSCHSTEIN2020102209} is the most comprehensive, and will be used as the primary model in the energy expenditure modeling and analyses in this paper, with contributions from Liu et al.~\cite{Liu2017} due to their empirical data evaluations. Samiei et al.~\cite{samiei2023distributed,samiei2024allocating} and Selje et al.~\cite{selje2024comprehensive} presented machine-learning-based approaches to estimating flight energy consumptions. However, these data-driven methods have demonstrated accurate energy consumption estimates only within the range covered by the data, while their predictive accuracy beyond this range remains unexplored. Calderon Ochoa et al.~\cite{calderon2024identifying} presented an approach to estimating key pamaremters, such as the drag coefficient, of a multicopter for flight energy consumption modeling.

The energy optimization approaches for UASs can be categorized into controller-based and algorithm-based methods. 
Controlled-based methods were evaluated in Okulski and Ławry{\'n}czuk~\cite{Okulski2022}, who sought to improve drone maneuverability, by comparing a standard Model Predictive Control (MPC) method, a PID controller, and an MPC controller with the native algorithm augmented to a trajectory-guessing algorithm. They found that the augmented MPC controller performed about 15.7\% better than the PID controller with regards to the attitude of the drone, which is the stability and control during flight, yet it consumed roughly 15.7\% more energy than the PID controller (3.50Wh to 4.05Wh). 
It is noted that Okulski and Ławry{\'n}czuk~\cite{Okulski2022} didn't optimize energy directly, instead stating that attitude stabilization would lead to overall energy efficiencies. 
Our proposition lies within the context of reducing energy consumption specifically.

Among algorithmic approaches to energy optimization, the work of Shivgan and Dong~\cite{Shivgan2020} utilizes a genetic algorithm to seek the minimum energy expenditure along a path. 
The genetic algorithm did prove to be more effective at optimizing the energy expenditure for waypoint path calculation versus a greedy algorithm, with a range of 9.6\% to 81\% reduced energy expenditure from 10 waypoints to 100 waypoints, respectively. 
Unfortunately, the Greedy algorithm has not been tested with real-time adjustments of the drone, but rather in path forecasting, and thus a different approach is required for this paper's intended application of energy optimization achieved with real-time path correction. In addition, the research doesn't have empirical backing to support the algorithm's performance against external factors, which is incredibly disruptive in the realm of aerial flight, and we suggest our controller will be able to account for these external factors. 

Path planning formulations for energy optimization have also shown to be viable solutions to the route optimization problem.
Morbidi et al~\cite{morbidi} determine the minimum energy or time trajectories for a drone to travel from a hover at a start location to a hover at an end location.
In order to solve the problem, an objective function is defined which formulates the energy consumption as a function of the angular acceleration of the motors.
Subsequently, for a given time or energy budget, the optimization of the alternative function is performed, i.e. energy optimization for a given time budget and vice versa. 
Kreciglowa et al~\cite{kreciglowa} provide a solution to the minimum energy path planning problem, by smoothing trajectories and ensuring minimum acceleration, jerk, or snap over the course of the routes.
The research concludes that minimizing snap over the course of the trajectory consumes up to 18.1\% and 12.7\% less energy than minimization of acceleration and jerk, respectively.

The motivation of this study is to maximize the utility of the limited energy that is available for a drone.
The majority of previous publications have either prioritized other variables (e.g. time, stability) of the drone or utilized other methods of analyzing energy optimization. 
Our model will use physics-based dynamic models of drones to formulate an energy consumption equation, as well as a mechanics-based model for instantaneous energy consumption of the rotors that serves as the real-time optimization feedback. The design has the intention of empirically testing the model with air corridors in mind, as they will be the likely future application of UAS in urban areas. 
%

The three main contributions of this paper are summarised as follows. The first contribution is the development of a new control strategy for real-time airspeed adjustment to minimize the energy consumption (in units of joules per meter) of a drone during its cruise phase. The proposed real-time airspeed regulator guarantees that the drone airspeed fast converges to the optimal value when the value of the parameters in the formulation of energy consumption changes or is unknown. The second contribution is the development of an MPC strategy to generate the control inputs in the form of the motors' angular acceleration that minimizes the energy consumption during the drone's take-off (ascending) and landing (descending) phases. This part of the contribution was an extension of the work in~\cite{blakesley2022minimum} with newly added aerodynamic drag force terms in the formulation. An analysis of the benefit of having the landing incentive term in the formulation is also presented.   The third contribution is an analysis of three prevailing equations for downwash coefficient approximation. Given various flight conditions, these equations utilize underlying assumptions to tweak how the downwash coefficient is calculated, and the effects of each equation are compared in this paper. The proposed airspeed controller and MPC planner were evaluated in a simulated environment based on a real-world scenario. The results demonstrate the effectiveness of the proposed control and planning strategies. 

The remaining of the paper is structured as follows. Section~\ref{sec:problem_statement} presents the preliminaries. Section~\ref{sec:airspeed_control} presents the real-time airspeed regulation for the cruise phase using a start-of-the-art energy consumption model. Section~\ref{sec:preflight-opt} presents the motion planning strategy for the takeoff and landing phases. Section~\ref{sec:Results} shows the simulation results and analysis. Section~\ref{sec:Conclusion} concludes the paper. 

\begin{figure}[h]
\centering
\includegraphics[width=0.8\textwidth]{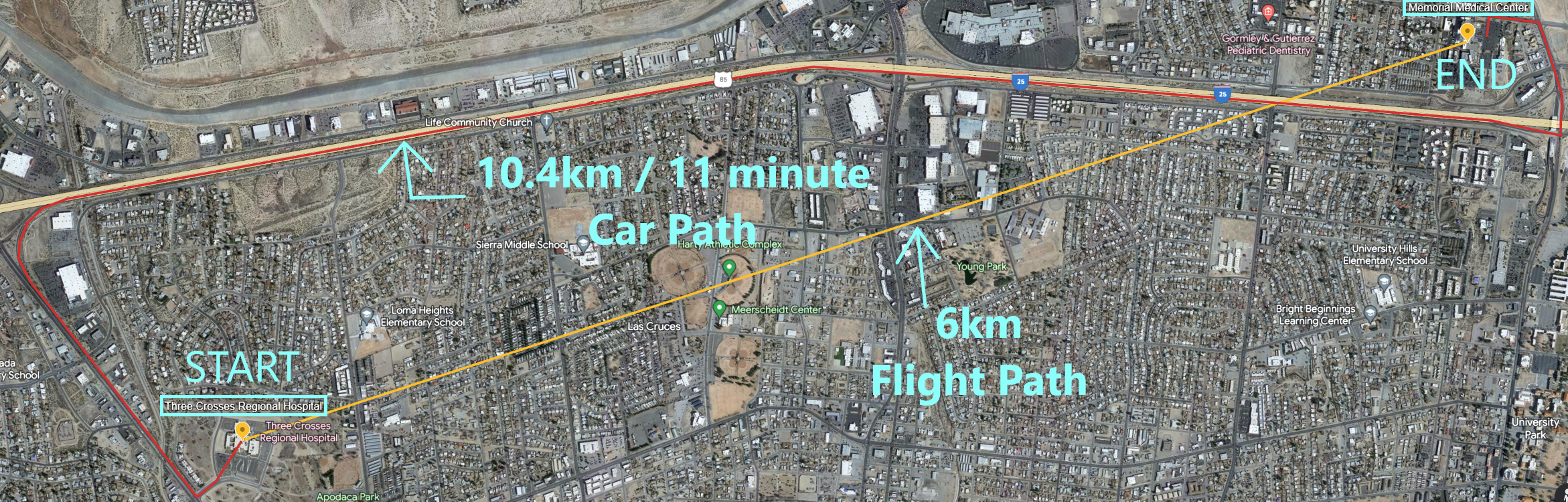}
\caption{Top-down view of the sample route between the Three Crosses Regional Hospital (left) and the Memorial Medical Center (right) in Las Cruces, NM, USA.}
\label{fig:Air Corridor between hospitals:top-down}
\end{figure}
\section{Preliminaries}
\label{sec:problem_statement}

\subsection{Background}

When working in urban environments, technologies that interact with or interface with public areas tend to follow a similar pattern: technology is limited by legislation. For Urban Air Mobility (UAM) or Advanced Air Mobility (AAM) applications, one of the greatest obstacles facing drone delivery services is safety regulations, and for good reasons. The Federal Aviation Administration (FAA) regulates the airspace around areas of interest, such as airports, hospitals with helipad usage, and military bases. Regulations have usually been established with private air travel in mind, but the stakeholders who develop or adopt UAM/AAM technologies need to coordinate with FAA to optimize the use of the national airspace. In recent years, FAA has introduced the idea of air corridors~\cite{faa_corridor} - passageways in the airspace above cities that would allow for different classes of air vehicles to travel in regulated airspace. These air corridors provide UAM/AAM technologies with access to previously restricted airspace while also keeping public areas relatively safe from accidents, failures, or other miscellaneous mishaps that could endanger the populace. 

The proposed formulation will be analyzed using an idea of an air corridor that provides UAS technologies with access to previously restricted airspace while also keeping public areas relatively safe from accidents, failures, or other miscellaneous mishaps that could endanger the populous. 
According to FAA documentation~\cite{faa_corridor}, UAM drones would have access to airspace about 400ft - 700ft above ground level (AGL), and will be given a takeoff and landing zone of unspecified size to reach the air corridor. 
Inside the air corridor, a specified speed limit will be established. 
Additionally, there will be areas of strictly restricted airspace, such as around airports, that may are likely to be defined as no-fly zones.

\begin{figure}[h]
\centering
\includegraphics[width=0.8\textwidth]{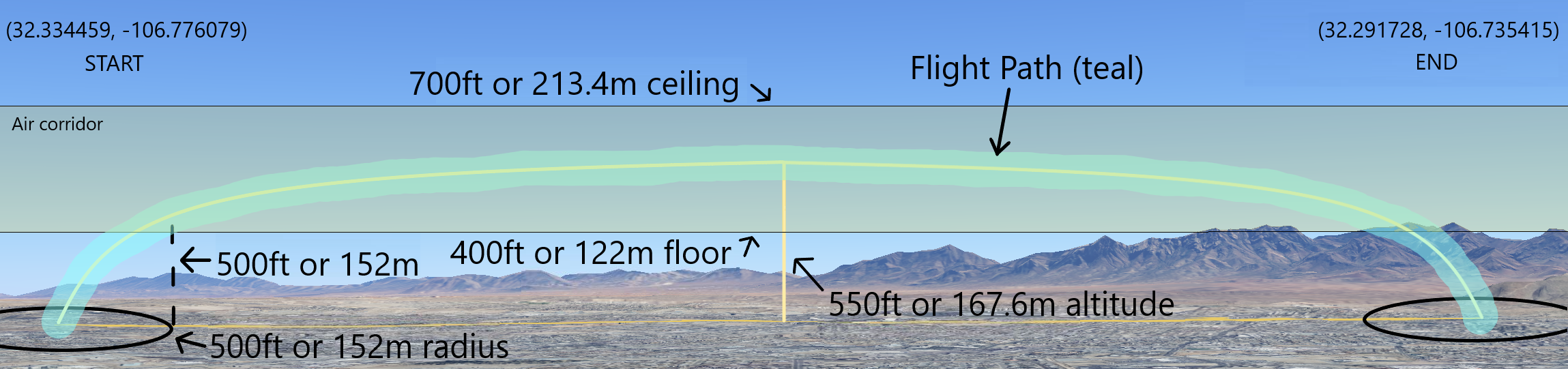}
\caption{A sample route through a conceptualized air corridor between two hospitals in Las Cruces, NM, USA.}
\label{fig:Air Corridor between hospitals:side-view}
\end{figure}

\subsection{Proposed Scenario}

In this paper, we consider the problem of minimizing the energy consumption for a package-delivery drone to transport a medical supply between predefined origin-destination pair.
An example scenario is shown in Fig.~\ref{fig:Air Corridor between hospitals:top-down}, where a medical package needs to be delivered from the Three Crosses Regional Hospital to the Memorial Medical Center in Las Cruces, NM, USA. The ongoing major construction of the I-25 freeway connecting the two locations would significantly slow down ground traffic. A delivery drone is proposed to do perform delivery by going through a predefined path shown in Fig.~\ref{fig:Air Corridor between hospitals:side-view}. The take-off and landing areas are selected to be 500 ft in radius, and extend up 500 ft into the air, while the air corridor itself will be between 400ft and 700ft AGL. 

\subsection{Problem Statement}

While time minimization is often critical in medical applications, it is assumed that a drone also needs to minimize its energy consumption to accommodate environmental uncertainties, such as wind gusts and turbulence, battery capacity degradation due to high temperature, etc. 
The entire flight path is divided into three phases: takeoff, cruise, and landing. In the cruise phase, the drone is assumed to fly inside the air corridor with a constant altitude and a fixed orientation. 
It is assumed that drone airspeed is the only variable that can be used for energy minimization (see Section~\ref{sec:airspeed_control}).  
In the takeoff and landing phases, the motor spinning speeds of a drone can be determined freely such that the resulting drone trajectory minimizes the energy consumption while satisfying the spatial constraints (e.g., starting and ending positions and corresponding velocities) and operational constraints (e.g., the limits on attitude angles and accelerations of the drone) (see Section~\ref{sec:preflight-opt}).  


\section{Realtime Energy-Minimization Airspeed Regulation in Cruise Phase\label{sec:airspeed_control}}

In this section, we propose a feedback control strategy to seek the airspeed command that minimizes the energy consumption of a drone in its cruise phase.  We first introduce the start-of-the-art energy expenditure model and subsequently outline the proposed feedback control method based on it.

\subsection{Energy per Meter Formulation}


Kirschstein~\cite{KIRSCHSTEIN2020102209} introduced a formulation for calculating energy expenditure over a distance, which Zhang et al.~\cite{zhang2021energy} standardized as energy per meter (EPM) and categorized Kirschstein's model as a component model based on helicopter operations. The core assumption of the component model is that the power consumed during hovering operation is approximately equivalent to the power consumption for takeoff, landing, and level flight operations. The equation for the thrust force during the cruise phase of a flight is given by
\begin{equation} 
  T= \sqrt{\left(g\sum_{k=1}^{3}m_{k}\right)^{2}+\left(\frac{1}{2}\rho\left(\sum_{k=1}^{3}C_{D_{k}}A_{k}\right)v_{a}^{2}\right)^{2}+\rho\left(\sum_{k=1}^{3}C_{D_{k}}A_{k}\right)v_{a}^{2}mg\sin\theta},\label{eqn:T}
\end{equation}
where index $k$ accounts for the three sections of the drone used for analysis: drone body, drone battery, and the payload attached to the drone, $\theta$ is the flight path angle, which is assumed to be zero in the paper, $v_{a}$ is the airspeed. Other parameters are assumed constant and are defined in Tables~\ref{tbl:indepent_Parameter_values} and~\ref{tbl:drone_parameters}.

The energy expenditure as a function of the airspeed is given by~\cite{KIRSCHSTEIN2020102209}
\begin{equation}
EPM=\frac{1}{\eta}\left(\frac{\kappa Tw}{v_{a}}+\frac{1}{2}\rho\left(\sum_{k=1}^{3}C_{D_{k}}A_{k}\right)v_{a}^{2}+\frac{\kappa_{2}\left(g\sum_{k=1}^{3}m_{k}\right)^{1.5}}{v_{a}}+\kappa_{3}\left(g\sum_{k=1}^{3}m_{k}\right)^{0.5}v_{a}\right)+\frac{P_{avio}}{\eta_{c}v_{a}} \label{EPM},
\end{equation}
where $\kappa$ is an up-scaling factor assumed to be 1.15 according to~\cite{McCormick1}, $\kappa_{2}$ is an experimental parameter, specified in Table~\ref{tbl:parameters_computed}, which is determined by propeller area and air density, $\kappa_{3}$ is an experimental parameter, specified in Table~\ref{tbl:parameters_computed}, which is a constant of proportionality between thrust and angular rotor speed, $w$ is the downwash coefficient, $\eta$ is the power transfer efficiency, $\eta_{c}$ is the battery charging efficiency, and $P_{avio}$ is the power required for the electronic avionic components.

\subsection{Downwash Coefficient Calculation}

The downwash coefficient $w$ in Eqn.~\eqref{EPM}, also referred to as the induced velocity, can be computed in different ways.  In this subsection, we introduce three different approaches to computing it, namely, the Root ($w_R$), Hover ($w_H$), and Glauert ($w_G$) approximations, respectively. 

The Root approximation of the downwash coefficient, $w_R$, is based on the following equation that was derived in~\cite{McCormick1}
\begin{equation}
w_R= \frac{T}{2n\rho\varsigma\sqrt{\left(\nu_{a}\cos\alpha\right)^{2}+\left(\nu_{a}\sin\alpha+w_R\right)^{2}}}, \label{downwash}
\end{equation}
and $w_R$ can then be computed by finding the real root of the following 4th-order polynomial equation
\begin{equation}
w_R^{4}+2w_R^{3}\nu_{a}\sin\alpha+w_R^{2}\nu_{a}-\left(\frac{T}{2n\rho\varsigma}\right)^{2}=0,\label{downwash_2}
\end{equation}
where $\alpha$ is the angle of attack and is given by
\begin{equation}
\alpha=\arctan\left(\frac{\frac{1}{2}\rho\left(\sum_{k=1}^{3}C_{D_{k}}A_{k}\right)\nu_{a}^{2}}{g\sum_{k=1}^{3}m_{k}}\right).\label{eqn:alpha}
\end{equation}
Finding the analytical roots of Eqn.~\eqref{downwash_2} is nontrivial. A common numerical approximation technique that involves the Newton-Raphson method \cite{ben1966newton} 
proves quite effective. 
The four roots usually include two complex roots with non-zero imaginary parts, a real negative root, and a real positive root, to which the real positive root is chosen as the solution for this work. 


The Hover approximation of the downwash coefficient, $w_H$, is given by~\cite{Leishman1}
\begin{equation}
w_{H}=\sqrt{\frac{T}{2\rho A}},\label{eqn:w_hover}
\end{equation}
which assumes that the hover operation is the most energy-taxing flight condition. While this approach is computationally efficient, it neglects the energy consumption of the horizontal flight, which can be significant if the airspeed is greater than a certain value. 


The Glauert approximation of the downwash coefficient, $w_G$, is based on Glauert's work in~\cite{glauert_1927}, which is given by
\begin{equation}
w_{G}=\frac{T}{2\rho A \nu_{a}}.\label{eqn:w_glauert}
\end{equation}
This approach relies on the phenomenon that helicopter blades will eventually behave like airfoils at high speeds and it produces an accurate approximation of the downwash coefficient for high airspeeds. 



\begin{figure}[h]
\centering
\includegraphics[width=0.7\textwidth]{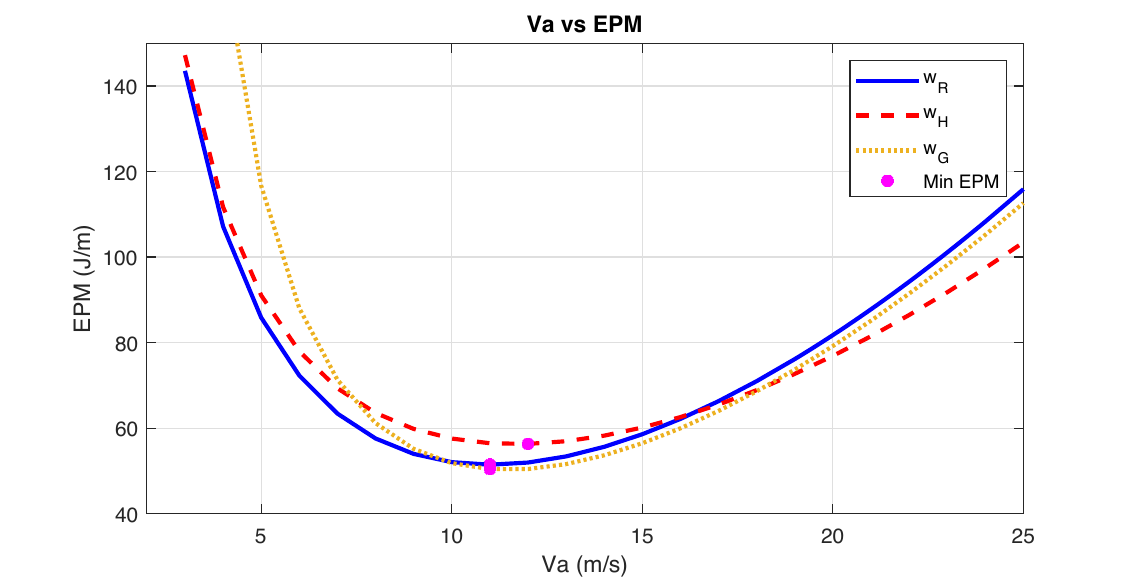}
\caption{EPM results using the Root ($w_R$), Hover ($W_H$), and Glauert ($W_G$) approximations for the downwash coefficient, respectively.}
\label{fig:3 induced velocity}
\end{figure}

Figure~\ref{fig:3 induced velocity} demonstrates the results of EPM using the Root, Hover, and Glauert approximations for the downwash coefficient, corresponding to Eqns.~\eqref{downwash_2},~\eqref{eqn:w_hover}, and~\eqref{eqn:w_glauert}, respectively.  Among these three approximations, the Root approximation is a purely physics-based derivation. Solving a 4th-order polynomial equation is very computationally costly compared to the other approximations, but it is the most theoretically accurate equation for forward, level flight. For multicopter, hovering is the most energy-intensive flight condition~\cite{Leishman1}, whose energy consumption dominates the energy consumption of the horizontal flight when the airspeed is relatively small. It can be seen in Fig.~\ref{fig:3 induced velocity} that the Root and Hover approximations generate close EPM results when the airspeed is less than $5\;\text{m/s}$ but start to deviate significantly when the airspeed is greater than $15\;\text{m/s}$. The Root and Glauert approximations closely match each other with only a small variance in magnitude when the airspeed is greater than $10\;\text{m/s}$. However, the Glauert approximation is certainly error-prone at low speeds when compared to the Root approximation.

\subsection{Energy Expenditure Calculations}
\begin{figure}[h]
\centering
\includegraphics[width=\linewidth]{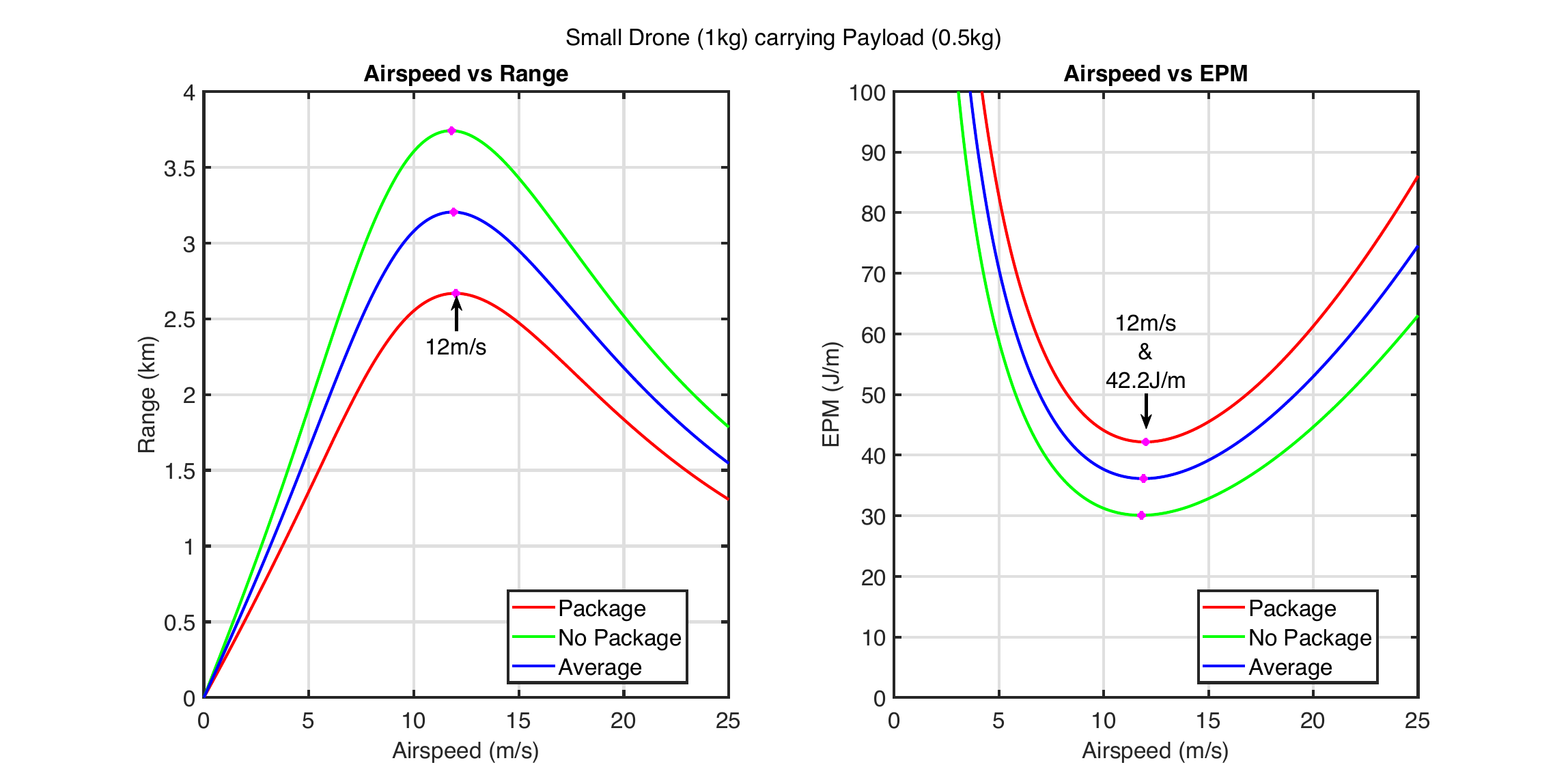}
\caption{Airspeed vs range (left) and Airspeed vs EPM (right) for a small drone with parameters in Table~\ref{tbl:drone_parameters}. }
\label{fig:EPM vs Velocity for small drone}
\end{figure}
Figure~\ref{fig:EPM vs Velocity for small drone} shows example curves of the airspeed vs EPM and airspeed vs range for a small delivery drone using the parameters in Table~\ref{tbl:drone_parameters}. Although the airspeed that minimizes the energy consumption during the cruise phase can be numerically obtained offline using Eqn.~\eqref{EPM}, it does not consider the disturbances (e.g., wind gusts) and parameter uncertainties in real-world applications. Therefore, we proposed a fast feedback control strategy to recursively adjust the airspeed command such that it converges to the optimal value to accommodate time-varying parameters in Eqn.~\eqref{EPM} such as wind. 

\subsection{Feedback Control Strategy for Airspeed Regulation in Cruise Phase}

Let $\overline{EPM}\left(V_a, k\right)$ be the EPM value at time step $k$, computed using the airspeed, $V_a$, and a set of nominal parameters, whose values can be inaccurate. Consider $\overline{EPM}\left(V_a+\Delta V, k\right)$, which is the EPM value computed using a perturbed airspeed, $V_a + \Delta V$, at time step $k$, where the perturbation is assumed to be positive for the proposed method, i.e., $\Delta V > 0$. Letting $\triangledown \overline{EPM}\left(V_a,k\right)$ be the gradient of $\overline{EPM}\left(V_a, k\right)$ with a small perturbation $\Delta V$ at time step $k$, we have
\begin{equation}
\triangledown \overline{EPM}\left(V_a, k\right)=\frac{\overline{EPM}\left(V_a+\Delta V, k\right)-\overline{EPM}\left(V_a, k\right)}{\Delta V}.
\label{EPM_grad}
\end{equation}
The airspeed command can be obtained 
\begin{equation}
V_{a}^{cmd}=V_{a}-\left(k_{p}\triangledown \overline{EPM}+k_{d}\frac{\text{d}\left(\triangledown \overline{EPM}\right)}{\text{d}t}+k_{i}\int\triangledown \overline{EPM}\right), \label{PID_controller}
\end{equation}
where $k_p$, $k_d$, and $k_i$ are positive constant control gains. 

In Eqn.~\eqref{EPM_grad}, a predicted EPM value with a positive perturbed airspeed, $\overline{EPM}\left(V_a+\Delta V, k\right)$, is computed to compare with the current EPM value.  If $\overline{EPM}\left(V_a+\Delta V, k\right)$ is greater (smaller) than $\overline{EPM}\left(V_a, k\right)$, $\triangledown \overline{EPM}\left(V_a, k\right)$ will be positive (negative), which implies that $V_{a}+\Delta V$ yields a greater (smaller) EPM value. In other words, if the slope of the EPM-$V_a$ curve in Fig.~\ref{fig:EPM vs Velocity for small drone} is positive (negative), then $V_a$ should decrease (increase).  This gradient descent approach is enhanced by using the PID controller in Eqn.~\eqref{PID_controller}.



\begin{figure}[t]
\centering
\includegraphics[width=0.6\columnwidth]{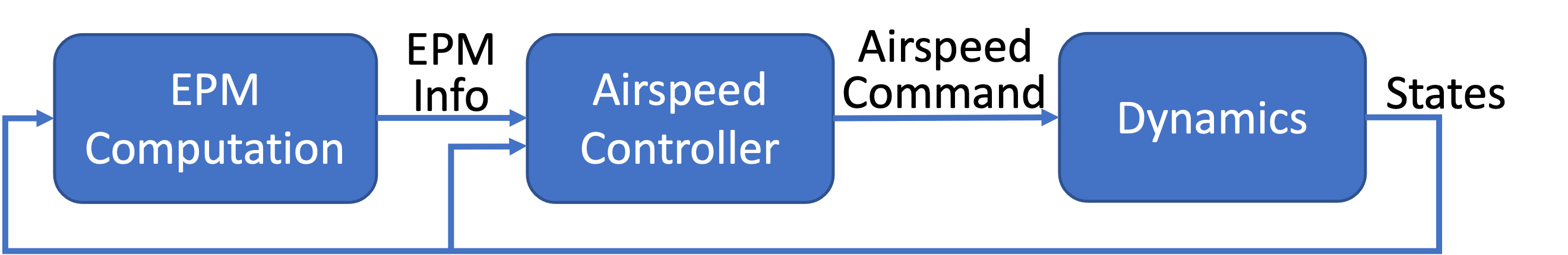}
\caption{\label{fig:airspeed-controller}Block diagram for the proposed realtime airspeed controller.}
\end{figure}

Figure~\ref{fig:airspeed-controller} shows the block diagram for the proposed airspeed strategy.  The "EPM Computation" function uses the system states that are needed in Eqn.~\eqref{EPM} to compute $EPM$ and its gradient $\triangledown EPM$. This EPM Info will be used by the airspeed controller in Eqn.~\eqref{PID_controller} to compute the airspeed command sent to the system dynamics.


Zhang et al.~\cite{zhang2021energy} observe that certain EPM equations, especially Kirschstein's formulation, are greatly affected by wind - leading to 100\%+ increases in EPM in some cases. Eqns. \eqref{EPM_grad} and \eqref{PID_controller} don't yet allow the controller to account for these wind disturbances, as the energy equations themselves don't account for wind. That is, the controller may reduce its velocity because energy demands will increase in the presence of wind, even at the same velocity, but the "optimal" velocity may not be accurate. Future improvements to the model intend on evaluating the energy demands from wind and then include these variables into the EPM equation for an all-encompassing model.

\section{Pre-flight Route Optimization for Takeoff and Landing Phases}\label{sec:preflight-opt}
\begin{figure}[t]
\centering
\includegraphics[width=\textwidth/2]{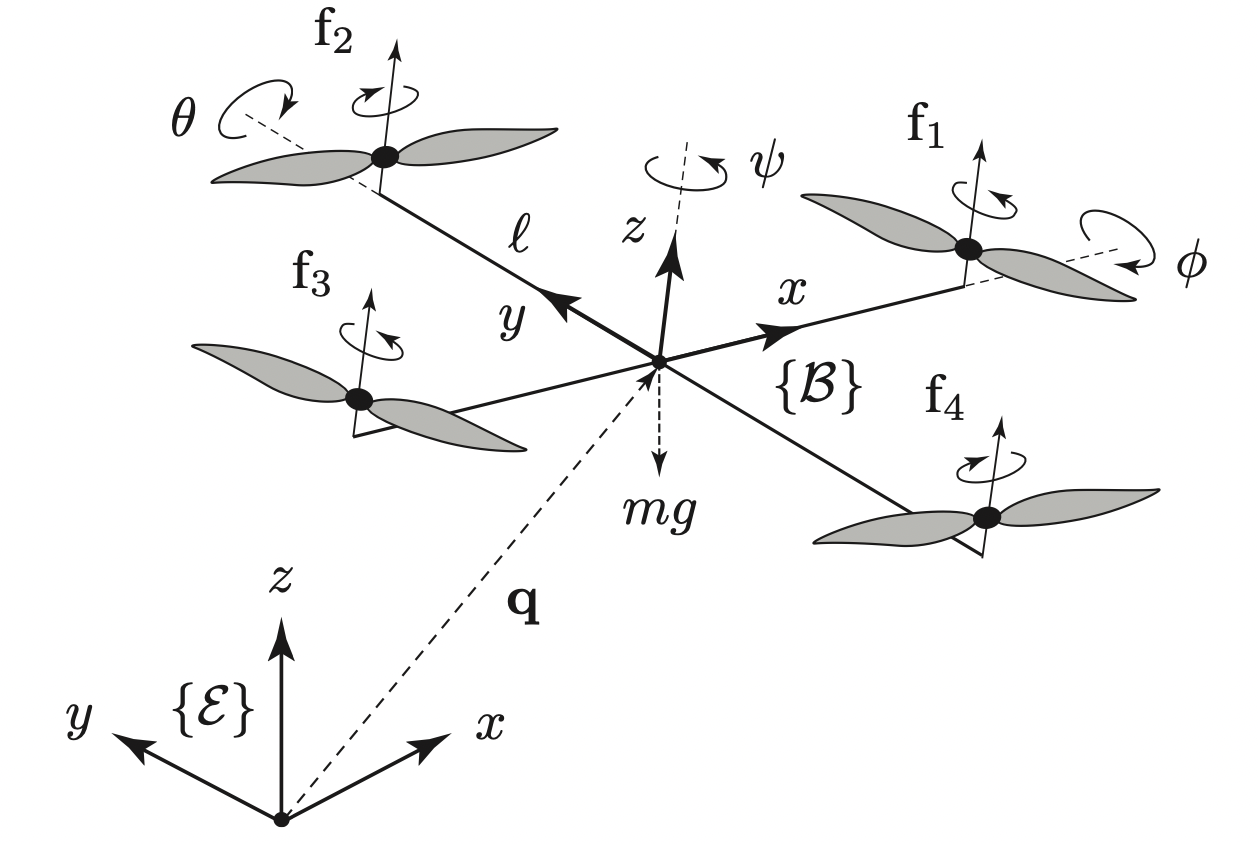}
\caption{The body ($\mathcal{B}$) and the earth ($\mathcal{E}$) coordinate frames; drone position ($x,y,z$) and attitude angles (roll ($\phi$), pitch ($\theta$), and yaw ($\psi$). Reprinted with permission from~\cite{morbidi} Copyright 2016 IEEE}
\label{fig:Dynamics}
\end{figure}

Let $(x,y,z)$ be the three-dimensional drone position in the Earth frame, $\mathcal{E}$; $\phi, \theta, \psi$ represent the roll, pitch, and yaw angles, and $\omega_i$, for $i = 1 ... 4$, denotes the rotational speed of motor $i$.
It is noted that motors 2 and 4 rotate in the clockwise direction, whereas 1 and 3 rotate anti-clockwise (see Figure~\ref{fig:Dynamics}). 
The state vector, $\bm{x}$, is defined as%
\begin{equation}
    \bm{x}(t) := [x(t), y(t), z(t), \dot{x} (t), \dot{y} (t), \dot{z} (t), \phi (t), \theta (t), \psi (t), \dot{\phi} (t), \dot{\theta} (t), \dot{\psi} (t), \omega_1 (t), \omega_2 (t), \omega_3 (t), \omega_4 (t)]^\top,
    \label{eq:State}
\end{equation}%
whilst the control input vector, $\bm{u}$, contains the angular acceleration of the four motors:
\begin{equation}
   \bm{u}(t) := [\alpha_1, \alpha_2, \alpha_3, \alpha_4]^\top,
    \label{eq:Control}
\end{equation}
where $\alpha_i = \dot{\omega}_i$.

\subsubsection{Quad-Rotor Dynamics}
The dynamics of the drone is based on fundamental helicopter dynamics.
It was first adapted to reflect a quad-rotor UAV in~\cite{bouabdallah} and is given by
\begin{equation} 
    \dot x = \frac{dx}{dt},
    \label{eq:EOM1}
\end{equation}
\begin{equation}
    \dot y = \frac{dy}{dt},
    \label{eq:EOM3}
\end{equation}
\begin{equation}
    \dot z = \frac{dz}{dt},
    \label{eq:EOM5}
\end{equation}
\begin{equation}
    \ddot{x} = \frac{T}{m} (\cos\phi \sin\theta \cos\psi + \sin\phi \sin\psi) - \frac{1}{2m} C_D \rho A_1 \dot{x} |\dot{x}|,
    \label{eq:EOM2}
\end{equation}
\begin{equation}
    \ddot{y} = \frac{T}{m} (\cos\phi \sin\theta \sin\psi - \sin\phi \cos\psi)- \frac{1}{2m} C_D \rho A_1 \dot{y} |\dot{y}|,
    \label{eq:EOM4}
\end{equation}
\begin{equation}
        \ddot{z} = \frac{T}{m} (\cos\phi \cos\theta) - g \left(\frac{2}{1+e^{-k ||\Delta \bm{x}||^2 }} - 1 \right)- \frac{1}{2m} C_D \rho A_1 \dot{z} |\dot{z}|,
    \label{eq:EOM6}
\end{equation}
\begin{equation}
    \dot \phi = \frac{d\phi}{dt},
    \label{eq:EOM7}
\end{equation}
\begin{equation}
    \dot \theta = \frac{d\theta}{dt},
    \label{eq:EOM9}
\end{equation}
\begin{equation}
    \dot \psi = \frac{d\psi}{dt},
    \label{eq:EOM11}
\end{equation}
\begin{equation}
    \ddot{\phi} = \frac{I_y - I_z}{I_x}\dot{\theta}\dot{\psi} + \frac{(F_2 - F_4)l}{I_x} - \frac{J \dot\theta \bar{\omega}}{I_x},
    \label{eq:EOM8}
\end{equation}
\begin{equation}
    \ddot{\theta} = \frac{I_z - I_x}{I_y}\dot{\phi}\dot{\psi} + \frac{(F_3 - F_1)l}{I_y} + \frac{J \dot\phi \bar{\omega}}{I_y},
    \label{eq:EOM10}
\end{equation}
\begin{equation}
    \ddot{\psi} = \frac{I_x - I_y}{I_z}\dot{\phi}\dot{\theta} + \frac{(M_1 - M_2 + M_3 - M_4)}{I_z},
    \label{eq:EOM12}
\end{equation}%
\begin{equation}
    \dot{\omega}_1 = \alpha_1,
    \label{eq:EOM14}
\end{equation}
\begin{equation}
    \dot{\omega}_2 = \alpha_2,
    \label{eq:EOM14}
\end{equation}
\begin{equation}
    \dot{\omega}_3 = \alpha_3,
    \label{eq:EOM14}
\end{equation}
\begin{equation}
    \dot{\omega}_4 = \alpha_4,
    \label{eq:EOM14}
\end{equation}
%
where parameter values used in this study follow~\cite{zhang2021energy} and~\cite{morbidi} and are given in Table~\ref{tbl:drone_parameters}.

It is noted that the gravitational term in Eqn.~\eqref{eq:EOM6} describing the acceleration in the vertical direction has been modified to incorporate a sigmoid-based function that multiplies gravitational acceleration, $g$, which depends on the squared distance to the destination, $||\Delta \bm{x}||^2 :=  (x_f - x)^2 + (y_f - y)^2 + (z_f - z)^2$, where $\bm{x}_f = [x_f, y_f, z_f]^\top$ is the final position described in Cartesian coordinates.
Once the drone approaches $\bm{x}_f$, the value of the sigmoid-based function converges to zero at a speed governed by the decay rate $k$.
The product of this modification with $g$ permits a drone to land at its final destination. 
This modification was first introduced in~\cite{blakesley2022minimum} and without it, the optimization time horizon $t_f$ dictates how long the drone is airborne, which can lead to energy sub-optimal solutions if  $t_f$ is set to a large value. 
This means that the value of $t_f$ could have a significant impact on the resulting trajectory if the landing incentivization is not included. 
More discussion and analysis on how the landing incentives effect the proposed framework and solution is provided in Section~\ref{sec:Results} and in~\cite{blakesley2022minimum}.

%
\subsubsection{Minimum Energy Route Optimization}
The instantaneous energy consumption, $E_i$, for any point in time is given by
\begin{equation}
    \begin{aligned}
        E_i = \; \frac{RT_f^2}{K_T^2} &  + \frac{T_f}{K_T}(\frac{2RD_f}{K_T} + K_T) \omega_i(t) \\
                                    & + \Big(\frac{D_f}{K_T}(\frac{RD_f}{K_T} + K_T) + \frac{2RT_f k_\tau}{K_T^2} \Big)  \omega_i^2(t) \\
                                    & + \frac{k_\tau}{K_T}(\frac{2RD_f}{K_T} + K_T) \omega_i^3(t) \\
                                    & + \frac{Rk_\tau^2}{K_T^2} \omega_i^4(t) \\
                                    & + \frac{RJ^2}{K_T^2} \dot\omega_i^2(t),
    \end{aligned}
    \label{eq:EnergyEQ}
\end{equation}
where all parameters are given by Table~\ref{tbl:parameters_computed}.

Furthermore, state vectors $\bm{x_0}$ and $\bm{x_f}$ are defined in order to specify the initial and final state respectively.
The objective function is formulated to reduce the energy consumption over the whole trajectory, such that:
\begin{equation}
    \begin{aligned}
        \min_{\omega_i} \sum_{i = 1}^{4} & \int_{t_0}^{t_f} E_i \; dt. \\
            s.t. \; \; \; \; \bm{x}(t_0) & = \bm{x_0}, \\
                             \bm{x}(t_f) & = \bm{x_f}, \\
                                  0 \leq & ~\omega_i \leq \omega_{max}, \\
                                  0 \leq & ~z, \\
                            -\pi/10 \leq & ~\phi \leq \pi/10, \\
                            -\pi/10 \leq & ~\theta \leq \pi/10 \\
                            \eqref{eq:EOM1}~-&~\eqref{eq:EOM12}
    \end{aligned}
    \label{eq:ObjectiveFunction}
\end{equation}
The optimization problem~\eqref{eq:ObjectiveFunction} is first discretised into 500 nodes using trapezoidal collocation method~\cite{kelly}, and is then solved numerically using the interior point line search filter method as implemented in the IPOPT solver~\cite{wachter}. 
The acceptable convergence tolerance and the maximum number of iterations were set to $10^{-4}$ and 5000 respectively.

\section{Simulation Results\label{sec:Results}}

\begin{figure}[t!]
\centering
\includegraphics[width=0.7\textwidth]{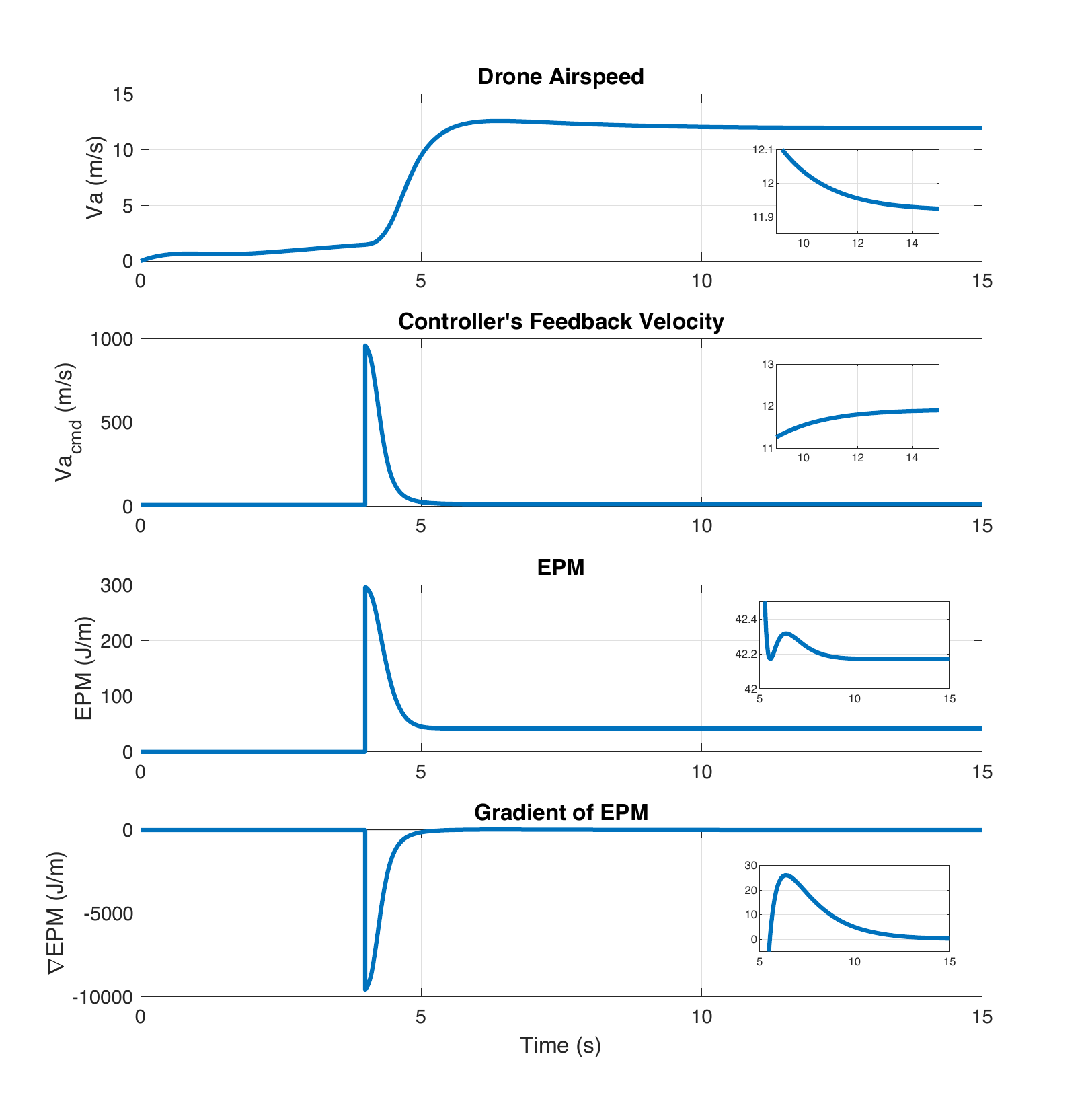}
\caption{Time evolution of $V_{a}$, EPM, $\triangledown$EPM, and $V_{a}^{cmd}$, where the airspeed controller was triggered at 4s. }
\label{fig:Scope of controller}
\end{figure}

\subsection{Result for Airspeed Regulation in Cruise Phase}

Recall that Fig.~\ref{fig:EPM vs Velocity for small drone} is generated by using Eqns.~\eqref{eqn:T} -~\eqref{eqn:alpha} with airspeed ranging from 0m/s to 25 m/s, and it serves as the theoretical baseline for the EPM equations from Zhang et al.~\cite{zhang2021energy}. Figure~\ref{fig:EPM vs Velocity for small drone} is used to validate the performance of the proposed controller in Eqn.~\eqref{PID_controller}. 
Figure~\ref{fig:Scope of controller} shows the time evolution of $V_{a}$, EPM, $\triangledown$EPM, and $V_{a}^{cmd}$, where the airspeed controller was triggered at 4 second. It can be seen that the proposed airspeed controller in Eqn.~\eqref{PID_controller} was able to recursively generate the airspeed command that converges to the optimal value that minimizes the EPM value. The input values of $\triangledown$EPM confirms the trendline of Airspeed ($V_{a}$) vs EPM in Fig.~\ref{fig:EPM vs Velocity for small drone}, as the decreasing EPM trendline would indicate a negative value of $\triangledown$EPM. 
A closer look at the $V_{a}$ and EPM curves shows where the optimal values of $V_{a}$ and EPM are located. These optimal values correspond to the calculated values in Fig. \ref{fig:EPM vs Velocity for small drone}, which are approximately 12 m/s for $V_{a}$ and 42.17 J/m for EPM, compared to the controller's values of 11.9 m/s and 42.17 J/m, thus providing credibility to the controller's accuracy. 



\subsection{Result for Route Optimization in Takeoff and Landing Phases}

\begin{figure}[h]
    \centering\includegraphics[width=1.02\textwidth]{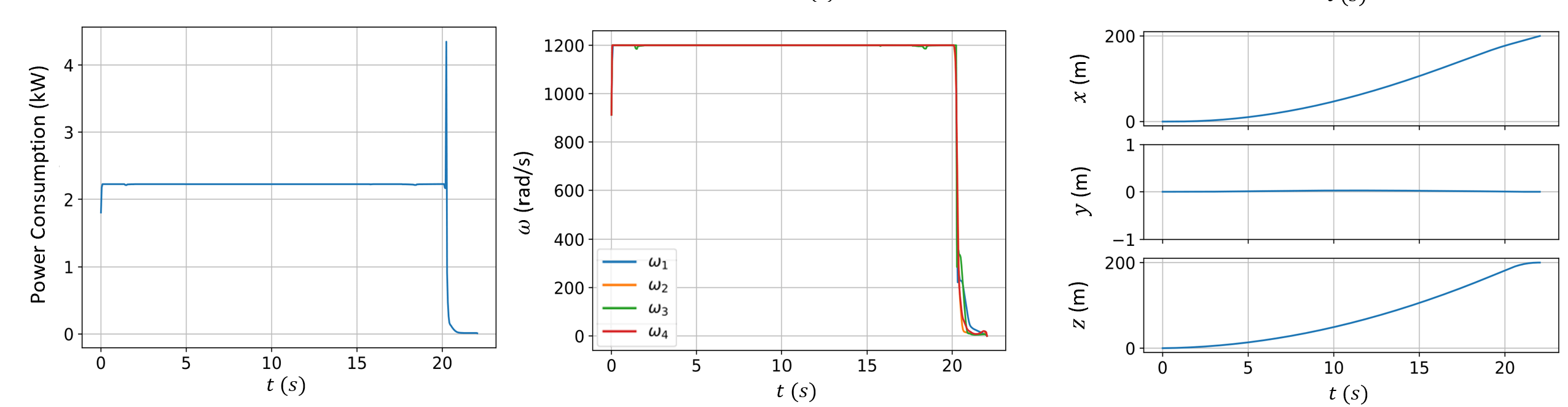}
    \caption{Take-off phase: energy consumption~(kJ) (left), angular velocity of the motors~(rad/s) (center), and spatial coordinates~(m) (right).}
    \label{fig:TakeOff_Results}
\end{figure}
\begin{figure}[h]
    \centering
    \includegraphics[width=1.02\textwidth]{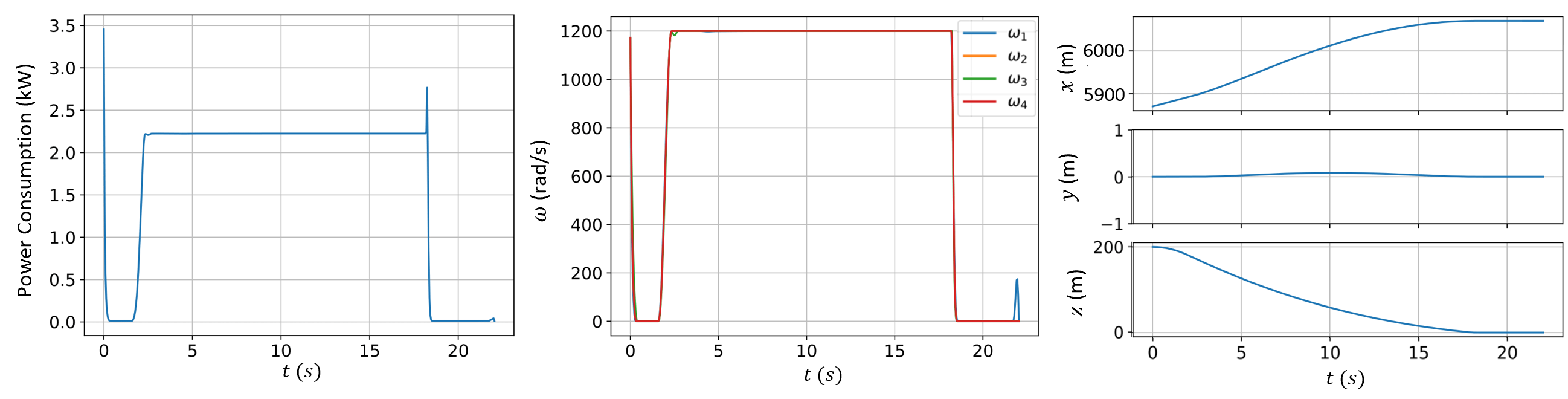}
    \caption{Landing phase: energy consumption~(kJ) (left), angular velocity of the motors~(rad/s) (center), and spatial coordinates~(m) (right).}
    \label{fig:Landing_Results}
\end{figure}
%

In the optimization of the takeoff stage, $\bm{x_0}$ and $\bm{x_f}$ in~\eqref{eq:ObjectiveFunction} are set to
\begin{equation}
    \bm{x_0} := [0 \times 12, 912.32~m/s \times 4 ]^\top,
    \label{eq:InstanceConstraints_t0}
\end{equation}
\begin{equation}
    \bm{x_f} := [200~m, 0, 200~m, 11~m/s, 0, 0, 0, 0.3~rad, 0 \times 4, 1172.3~m/s \times 4 ]^\top,
    \label{eq:InstanceConstraints_tf}
\end{equation}
and $t_f = 22$~s.
Figure~\ref{fig:TakeOff_Results} shows the resulting power consumption, the angular velocity of four motors, and spatial coordinates of the optimal trajectory of the UAV during the take-off stage. 
The power consumption is constant throughout the majority of the time horizon where the motors operate at their maximum rotational speed.
The energy consumption for this maneuver is 45.12~kJ.
In order to solve this problem, the term multiplying $g$ in Eqn.~\eqref{eq:EOM6} was set to 1 (equivalent of $k \rightarrow \infty$) in order to neutralize the impact of landing incentivization.
It is noted that the modification introduced in~\eqref{eq:EOM6}  only works effectively if $\dot{x}(t_f) =\dot{y}(t_f) = \dot{z}(t_f)= 0$, otherwise the UAV will be forced to escape the region of low gravity, which will lead to an increase in energy consumption.
This is the case here where $\dot{x}(t_f) = 11$~m/s.

\begin{table}[h!]
  \caption{Parameter values independent of drone~\cite{zhang2021energy,morbidi}*}
  \label{tbl:indepent_Parameter_values}
  \begin{tabularx}{\linewidth}{l*{6}{Y}}
    \toprule
               Variable & Symbol & Value \\[0pt]
    \midrule
    Air density [$kg/m^{3}$] & $\rho$ & 1.225\\
    Acceleration due to gravity [$m/s^{2}$] &  g &  9.807\\
    Battery power transfer efficiency (from battery to propeller) & $\eta$ & 0.7\\
    Up-scaling factor & $\kappa$ & 1.15\\
    Specific energy of battery [$J/kg$] & $s_{batt}$ & 540,000\\
    Safety factor of battery & $f$ & 1.2\\
    Depth of charge of battery & $\gamma$ & 0.5\\
    \bottomrule
  \end{tabularx}
  *Reprinted with permission from~\cite{zhang2021energy} Copyright 2020 Elsevier Ltd. and~\cite{morbidi} Copyright 2016 IEEE.
\end{table}
\begin{table}[h!]
    \caption{Key drone parameters~\cite{zhang2021energy,morbidi}*}\label{tbl:drone_parameters}
  \begin{tabularx}{\linewidth}{l*{6}{Y}}
    \toprule
              Variable & Symbol & Drone\\[0pt]
    \midrule
    Number of rotors & $n$ & 4\\
    Number of blades in one rotor & N & 4\\
    Offset between blade root and motor hub & $\epsilon$ & 0.004\\
    Mass of a blade (included in $m_1$) [$kg$] & $m_b$ & 0.0055\\
    Radius of the propeller [$m$] & $r$ & 0.127\\
    Spinning area of one rotor [$m^{2}$] & $\varsigma$ & 0.0507\\
    Distance between rotor and drone center of mass [$m$] & $l$ & 0.175\\
    Mass of drone body [$kg$] & $m_{1}$ & 1.07\\
    Mass of battery [$kg$] & $m_{2}$ & 1\\
    Mass of payload [$kg$] & $m_{3}$ & 0.5\\
    Drag coefficient of drone body & $C_{D_{1}}$ & 1.49\\
    Drag coefficient of battery & $C_{D_{2}}$ & 1\\
    Drag coefficient of payload & $C_{D_{3}}$ & 2.2\\
    Thrust coefficient of the propeller & $C_T$ &  0.0048\\
    Torque coefficient of the propeller & $C_Q$ &  0.00023515\\
    Projected area of drone body [$m^{2}$] & $A_{1}$ & 0.0599\\
    Projected area of battery [$m^{2}$] & $A_{2}$ & 0.0037\\
    Projected area of payload [$m^{2}$] & $A_{3}$ & 0.0135\\
    Profile power factor [$(m/kg)^{1/2}$] & $\kappa_{2}$ & 0.790\\
    Thrust - Rotor speed scaling factor [$(m/kg)^{-1/2}$] & $\kappa_{3}$ & 0.0042\\
    Motor moment of inertia [$kg m^2$] & $J_m$ & 4.9 $\times$ $10^{-6}$\\
    X component of body inertia [$kg m^2$] & $I_x$ & 0.081\\
    Y component of body inertia [$kg m^2$] & $I_y$ & 0.081\\
    Z component of body inertia [$kg m^2$] & $I_z$ & 0.142\\
    Decay rate of landing incentive & $k$ & 3\\
    Flight angle of drone & $\theta$ & $-$\\
    Torque constant of the motor[$Vs/rad$] & $K_T$ & 0.01038\\
    Motor friction torque [$Nm$] & $T_f$ & 4 $\times$ $10^{-2}$\\
    Resistance in the phase winding [$\Omega$] & $R$ & 0.2\\
    Viscous damping coefficient of the motor [$Nm s/rad$] & $D_f$ & 2 $\times$ $10^{-4}$\\
    Maximum angular velocity of the motor [$rad/s$] & $\omega_{max}$ & 1200 \\
    \bottomrule
    \end{tabularx}    
    *Reprinted with permission from~\cite{zhang2021energy} Copyright 2020 Elsevier Ltd. and~\cite{morbidi} Copyright 2016 IEEE.
\end{table}


\begin{table}[t!]
    \caption{Parameter values requiring calculation~\cite{zhang2021energy,morbidi}*}
    \label{tbl:parameters_computed}
    \begin{tabularx}{\textwidth}{XcY}
    \toprule
               Variable & Symbol & Equation \\[0pt]
    \midrule
    ~\cite{KIRSCHSTEIN2020102209} Downwash coefficient & $w$ & $\frac{T}{2n\rho\varsigma}\,=\,w\sqrt{\left(w-{v_{a}}\sin\alpha\right)^{2}+\left({v_{a}}\cos\alpha\right)^{2}}$\\ 
    & & $v_{a}$ - drone airspeed; $\alpha$ - angle of attack \\
    ~\cite{KIRSCHSTEIN2020102209} Angle of attack & $\alpha$ & $\alpha=\arctan\left(\frac{\frac{1}{2}\rho\left(\sum_{k=1}^{3}C_{D_{k}}A_{k}\right)\nu_{a}^{2}}{g\sum_{k=1}^{3}m_{k}}\right)$\\
    ~\cite{Liu2017} Profile power factor & $\kappa_{2}$ & $\sqrt{2\rho A}$\\
    ~\cite{Liu2017} Thrust - Rotor speed scaling factor & $\kappa_{3}$ & $T_{i} = \kappa_{3}\Omega_{i}^{2}$\\
    & & $i$ - individual rotor number\\    
    ~Overall mass of the UAV body & $m$ & $m_1$ + $m_2$ + $m_3$\\
    ~\cite{morbidi} Thrust factor & $k_b$ & $C_T \rho A r^2$\\
    ~\cite{morbidi} Drag factor & $k_\tau$ & $C_Q \rho A r^3$\\
    ~\cite{morbidi} Force Produced by the $i^{th}$ motor & $F_i$ & $k_b (\omega_i^2)$\\
    ~Resultant Thrust & $T$ & $\sum_{i = 1}^{4} F_i $\\
    ~\cite{morbidi} Moment Produced by the $i^{th}$ motor & $M_i$ & $k_\tau (\omega_i^2)$\\
    ~\cite{morbidi} Angular velocity component & $\bar{\omega}$ & $\omega_1-\omega_2+\omega_3- \omega_4$\\
    ~\cite{morbidi} Inertial moment & $J$ & $J_m + J_l$\\
    ~\cite{morbidi} Load moment of inertia & $J_l$ & $\frac{1}{4} n_B m_B(r - e)^2$\\
    ~Area covered by the propeller & $A$ & $\pi r^2$\\
    \bottomrule
  \end{tabularx}
  \endgraf 
  *Reprinted with permission from~\cite{zhang2021energy} Copyright 2020 Elsevier Ltd. and~\cite{morbidi} Copyright 2016 IEEE.
\end{table}


\vspace{10em}
Similarly to the takeoff stage, the trajectory optimization during the landing is conducted following the formulation introduced in Section~\ref{sec:preflight-opt} with $\bm{x_0}$ and $\bm{x_f}$ given by
\begin{equation}
    \bm{x_0} := [5870.21~m, 0, 200~m, 11~m/s, 0, 0, 0, 0.3~rad, 0 \times 4, 1172.3~m/s \times 4 ]^\top,
\end{equation}
\begin{equation}
    \bm{x_f} := [6070.21~m, 0 \times 15]^\top.
    \label{eq:InstanceConstraints_tf}
\end{equation}
Here, $t_f=22s$ and landing incentivization encourages the drone to land when it is the most energy optimal.
The solution to Eqn.~\eqref{eq:ObjectiveFunction} is shown in Fig.~\ref{fig:Landing_Results}.
It can be seen that the motors operate below the maximum angular velocity causing the UAV to descend.
This leads to non-constant power consumption and results in total energy usage of 36.47~kJ.

In order to commence the descent, the motor angular velocity is reduced to zero for 2~s, resulting in a brief free fall.
This poses no issues in the absence of external factors such as wind but should be considered in future test cases where such factors could affect the UAV's trajectory or its stability.
It should be noted that landing incentivization causes the UAV to reach the final destination at 16s (smaller than $t_f$).
After landing, it stays stationary for the remainder of the allotted time (see Fig.~\ref{fig:Landing_Results}).
\subsection{Benefit of Landing Incentivization}
\begin{figure}[h]
    \centering
    \includegraphics[width=\textwidth/2]{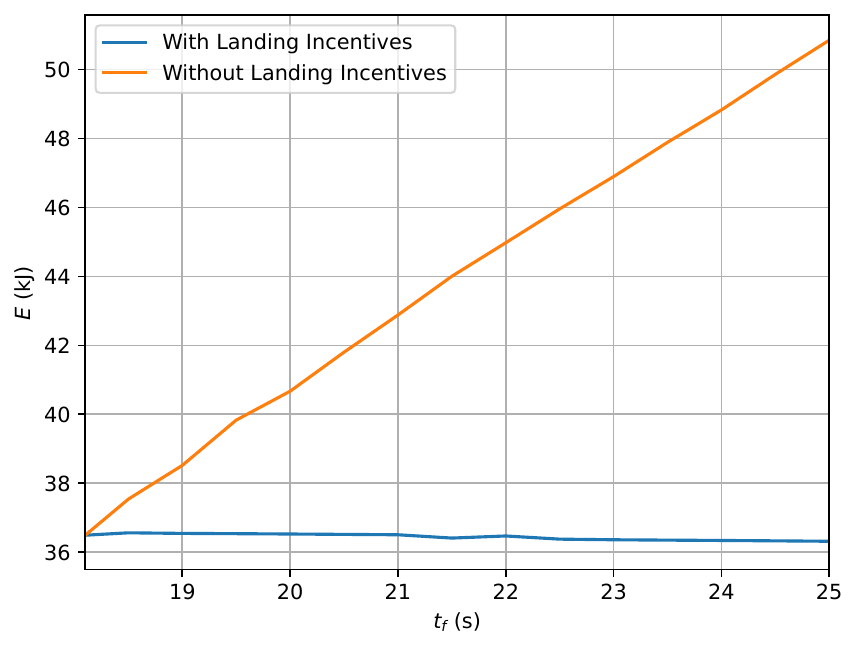}
    \caption{Comparison of the energy consumed in the landing phase using the models with and without landing incentives activated, respectively.}
    \label{fig:EnergyComp}
\end{figure}

Figure \ref{fig:EnergyComp} illustrates the energy consumption obtained for different values of $t_f$ with and without the landing incentive for the landing section of the flight.
The equation of motion without the landing incentive is given when~Eqn.\eqref{eq:EOM6} is replaced by the standard equilibrium equation of horizontal forces in the form of 
\begin{equation}
        \ddot{z} = \frac{T}{m} (\cos\phi \cos\theta) - g - \frac{1}{2m} C_D \rho A_1 \dot{z} |\dot{z}|.
    \label{eq:EOM6.1}
\end{equation}
Figure \ref{fig:EnergyComp} shows that in the absence of landing incentives, the energy consumption increases linearly with $t_f$, because the drone is unable to land and so travels increasingly slowly towards its destination whilst consuming more energy to counteract gravitational acceleration.
The optimal value to use for $t_f$ is very difficult to calculate and it can be seen that without the landing incentives, the value for $t_f$ can have a dramatic impact on the energy consumption of the UAV over the trajectory.
In order to find the true optimal trajectory without the landing incentives, the problem needs to solved for a range of selected $t_f$ to obtain the orange curve in Fig.~\ref{fig:EnergyComp}, which is very computationally inefficient. 
The blue curve, which represents the results when the landing incentives are included, shows that the optimal energy consumption does not depend on the selected $t_f$, which removes the impact of $t_f$ on the optimal result. 
The minimum flight time of the drone between the two locations is 16 seconds, which also corresponds to the minimum energy consumption.
It is important to note that if an unreasonably small value of $t_f$ is selected, the problem becomes unfeasible and the optimization solver will be unable to solve the problem, hence, Figure~\ref{fig:EnergyComp} only shows results for $t_f > 16$s.

The equation of motion defined in Eqn.~\eqref{eq:EOM6} that includes the landing incentives was introduced in~\cite{blakesley2022minimum}, in which analysis was done comparing the effects of the framework with and without the landing incentives against  a benchmark result in~\cite{morbidi}. The result shows that the application of the landing incentives has reduced travel energy consumption of a UAV route by 80\% compared to the result in~\cite{morbidi}. 
This paper has extended our previous work in~\cite{blakesley2022minimum} by adding the aerodynamic drag force terms in the formulation and using a set of parameters for a new drone, which is part of the novel contribution of this work.

\section{Conclusion\label{sec:Conclusion}}
In this paper, we explored the strategies for minimizing the energy consumption of the take-off, cruise, and landing flight modes of a multicopter in a simulated package-delivery mission. The proposed new feedback regulator is able to recursively generate airspeed commands and converges fast to the optimal airspeed that minimizes the energy consumption of a multicomputer in the cruise phase. The proposed model-predictive-control (MPC) approach is able to find the optimal motor acceleration profile in both the takeoff and landing phases. The benefit of the proposed landing incentive term in the MPC formulation was revealed and analyzed in eliminating the high impact of the selected optimization time horizon on the MPC solution.  


\section*{Acknowledgement}
Part of this work is supported by a New Mexico NASA EPSCoR Research Infrastructure Development (RID) grant, under the US National Aeronautics and Space Administration (NASA) Cooperative Agreement No.~NM-80NSSC19M0181. Part of this work is financially supported by the Engineering and Physical Sciences Research Council (grant number: EP/V002619/1) and the Department of Civil, Environmental, and Geomatic Engineering, University College London, UK.

\bibliography{ref}
\end{document}